\documentclass[useAMS,usenatbib]{mn2e}
\usepackage{graphicx}

   \title[VGS31b: a highly inclined ring along a filament in a void. Implication for the cold accretion.]{VGS31b: a highly inclined ring along a filament in a void. Implication for the cold accretion.}
  \author[M. Spavone et al.]{M. Spavone$^{1}$\thanks{E-mail: spavone@na.astro.it (MS)}, E. Iodice$^{2}$\\
     $^{1}$Dipartimento di Fisica e Astronomia, Universit\'a di Padova, Vicolo dell'Osservatorio 2, I-35122 Padova, Italy\\
    $^{2}$INAF-Astronomical Observatory of Naples, via Moiariello 16,
   I-80131 Napoli, Italy\\
        }
\begin{document}
\bibliographystyle{mn2e}
   \date{Accepted 2013 July 5.  Received 2013 July 1; in original form 2013 May 21}

\pagerange{\pageref{firstpage}--\pageref{lastpage}} \pubyear{2012}

\maketitle

\label{firstpage}

\begin{abstract}

VGS31b is a highly-inclined ring galaxy found along a filament in a void
\citep{Kreckel12}. Detailed photometry, by using {\it u, g, r, i, z}
SDSS images, has shown that the overall morphology of VGS31b is very
tricky, due to {\it i)} the presence of a highly inclined
($72^{\circ}$) ring-like structure, which reaches the galaxy center
tracing a ``spiral-like'' pattern, {\it ii)} a one sided tail towards
North-East and {\it iii)} a bar in the central regions
\citep{Beygu13}. Such structure is reasonably the result of a
``second event'' in the evolution history of this galaxy, which could
be a gravitational interaction with a companion galaxy or with the
environment.

The main aim of the present work is to address the most reliable formation
scenario for this object, by comparing the observed properties,
i.e. structure, baryonic mass, kinematics and chemical abundances,
with the theoretical predictions. In particular, we have used archival
spectroscopic data, to derive the metallicity in the ring: we found
a very low, sub-solar average value of $Z = 0.3 Z_{\odot}$, comparable
with other polar ring/disk galaxies, but lower than those measured for
ordinary spirals of similar luminosity. The study of the chemical
abundances in polar ring/disk galaxies and related objects has
received an increasing attention in recent years, since it has
revealed to be a key-parameter to disentangle among the formation
scenarios suggested for this class of objects: major merging, tidal
accretion or cold accretion.  In the present work we check the cold
accretion of gas through a ``cosmic filament'' as a possible scenario
for the formation of the ring-like structure in VGS31b.

\end{abstract}

%________________________________________________________________

\section{Introduction} \label{intro}
The formation and evolution of galaxies is one of the leading themes
of the modern observational cosmology. There are three main scenarios
proposed for galaxy formation: 1) the monolithic collapse
\citep{Egg62}, in which the collapse of the initial gas clouds quickly
form stars leading to the formation of spheroidal systems, while the
disks form later on through gas accretion; 2) the hierarchical
scenario, in which disks are the first to form and then interactions
and mergers lead to the formation of spheroids; 3) the external gas
accretion from filaments \citep{Com11}, in which disks form first and
may evolve without merger with other galaxies, forming spheroids at
the center through secular evolution acquiring gas from cosmic web
filaments.

The advent of new all-sky surveys and high resolution data, covering a
wide wavelength range, have strongly confirmed that gravitational
interactions and mergers affect the morphology and dynamics of
galaxies at all redshifts. In this framework, the study of peculiar
and interacting galaxies can shed light on the main processes at work
during galaxy interactions and on the influence of their
environment. However, how galaxies acquire their gas is still an open
issue in the models of galaxy formation. Recent theoretical works,
which are supported by many numerical simulations, have argued that
cold accretion plays a major role (\citealt{Kat93}; \citealt{Kat94};
\citealt{Ker05}; \citealt{Dek06}; \citealt{Dek08};
\citealt{Bou09}). \citet{Ker05} studied the physics of the {\it cold
  mode} of gas accretion in detail and they find that it is generally
directed along filaments, allowing galaxies to draw gas from large
distances. In particular, the cold accretion is a key mechanism for
providing gas to disk galaxies \citep{Bro09}.

Recent simulations of disk formation in a cosmological context,
performed by \cite{Age09}, revealed that the so-called chain galaxies
and clump clusters, found only at higher redshifts \citep{Elm07}, are
a natural outcome both of early epoch enhanced gas accretion from cold
dense streams and of tidally and ram-pressured stripped material
from minor mergers and satellites. This freshly accreted cold gas
settles into large disk-like systems. This scenario reproduces the
observed morphology and global rotation of disks and predicts both a realistic
metallicity gradient and a star formation rate (SFR) of $20
M_{\odot}/yr$. \cite{Age09} find solar metallicity for the inner
disk, while in the clump forming region it is only $\sim 1/10
Z_{\odot}$, owing to the accretion of pristine gas in the cold streams
mixed with stripped satellite gas.

Simulations also show that the interaction region between the newly
formed disk and the cold streams can also cause it to be misaligned
with the initial galactic disk. Based on very limited statistics,
\cite{Age09} suggest that this misalignment might not be typical, and
that it comes from a third cold stream that is perpendicular to the
main filament. More recent analysis shows that the accretion of gas
along misaligned filaments with respect to the disk plane are more
common, and it leaves traces down to low redshift (\citealt{Dek09N};
\citealt{Ros10}). An almost polar structure can result even as an extreme
case of this process and, as suggested by \cite{Age09}, it could be
responsible for forming polar disks.

Hydrodynamical simulations performed by \cite{Mac06} and \cite{Bro08}
have shown that the formation of a polar disk galaxy can occur
naturally in a hierarchical universe, where most low-mass galaxies are
assembled through the accretion of cold gas infalling along
filamentary structures. According to \cite{Mac06}, the polar disk
forms from cold gas that flows along the extended $\sim 1 Mpc$
filament into the virialized dark matter halo. The gas streams into
the center of the halo on an orbit that is offset from radial
infall. As it reaches the center, it impacts with gas in the halo of
the host galaxy and with the warm gas flowing along the opposite
filament. Only the gas accreted perpendicular to the major axis of the
potential can survive for more than a few dynamical lifetimes.

\cite{Bro08} argue that polar disk galaxies are extreme examples of
the misalignment of angular momentum that occurs during the
hierarchical structure formation: an inner disk starts forming shortly
after the last major merger at $z \sim 2$. Because of its gas rich
nature, the galaxy rapidly forms a new disk whose angular momentum is
determined by the merger orbital parameters. Later, gas continues to
be accreted but in a plane that is almost perpendicular to the inner
disk. At $z \sim 0.8$ the central galaxy is still forming stars in a
disk, while the bulk of new star formation is in a highly inclined
polar disk. The inner disk has exhausted its gas by $z\sim 0.5$, while
gas continues to fall onto the polar disk. From this point on, star
formation occurs exclusively in the polar disk, which remains stable
for at least 3 Gyrs. The formation mechanisms described above can
self-consistently explain both the morphology and kinematics of a polar
disk galaxy.

The observed similarities between polar ring/disk galaxies,
i.e. structure, gas content, age, \citep{Arn97, Iod02, Cox06, Spav10,
  Spav11} and similar formation history with disk galaxies, yield in
the latest decade, a renewed effort in studying the morphology and
kinematics of Polar Ring/Disk Galaxies (PRGs). These systems are
  made by a central spheroidal component and a polar structure of gas,
stars and dust \citep{Whi90}: the
existence of two orthogonal components of the angular momentum is a
consequence of a second event in their formation history. They can
thus be considered an ideal laboratory to study the physics of
accretion/interaction mechanisms, the disk formation and the dark halo
shape.

Recent studies on the prototype of PRGs, NGC4650A, revealed that the
polar structure in this object has disk-like morphology and
kinematics, rather than the nature of a ring (see \citealt{Arn97};
\citealt{Iod02}; \citealt{Gal02}; \citealt{Swa03}). Since then,
theoretical predictions have been re-addressed to understand how
different kinds of galaxy-galaxy and galaxy-environment interactions
lead to different morphologies and kinematics of PRGs and in
particular, if galaxies with ``pure rings'' could form with the same
mechanism that let to the grow of a polar disk.\\ Currently, in
addition to the cold accretion scenario widely described above, the
other two main formation processes proposed for accounting the PRGs
manifold are: {\it i)} a major dissipative merger and the {\it ii)}
tidal accretion of material (gas and/or stars) by a donor. In the
merging scenario, the PRG results from a ``polar'' merger of two disk
galaxies with unequal mass (\citealt{Bek98a}; \citealt{Bou05}). In the
accretion scenario, the polar ring/disk may form by a) the disruption
of a dwarf companion galaxy orbiting around an early-type galaxy
(ETG), or by b) the gas accretion on an ETG from the outskirts of a
tidally stripped disk galaxy, during a parabolic encounter
(\citealt{Res97}; \citealt{Bou03}; \citealt{Han09}).

Recent observational studies on PRGs \citep{Iod06,Spav10,Spav11}
single out the critical physical parameters that allow to disentangle
among the formation scenarios. They are 1) the total baryonic mass of
the polar structure vs.  that of central spheroid; 2) the kinematics
along both the equatorial and meridian planes; 3) the metallicity \&
SFR in the polar structure. By studying the chemical abundances in the
polar structure of three polar ring/disk galaxies, NGC4650A, UGC7576
and UGC9796, \citet{Spav10,Spav11} have led the way in implementing a
test for the cold accretion, tracing the formation history of these
objects by accounting for all the three parameters mentioned above. In
particular, the low, sub-solar metallicities derived for these objects
turn to be consistent with the estimates for a simulated polar disk
galaxy formed via cold accretion \citep{Snaith12}.

In this work, we would like to perform the same kind of analysis on
the galaxy VGS31b, which is a highly-inclined ring galaxy found along a
filament in a void \citep{Kreckel12}. Main aim of the present work is
to address the most reliable formation scenario for this object, by
comparing the observed properties, i.e. structure, baryonic mass,
kinematics and chemical abundances, with the theoretical predictions.

The paper has the following structure: in Sec.\ref{vgs31} are
described the main properties of VGS31b, in Sec.\ref{obs} we describe
the data reduction and analysis, in Sec.\ref{morph} and \ref{phot} we
show the results of the photometric analysis. In Sec.\ref{oxy},
\ref{ML} and \ref{SFR} we describe the study of the chemical
abundances of VGS31b, while discussion and conclusions are illustrated
in Sec.\ref{conc}.

\section{Properties of VGS31}\label{vgs31}

\begin{figure*}
\centering
\includegraphics[width=10cm]{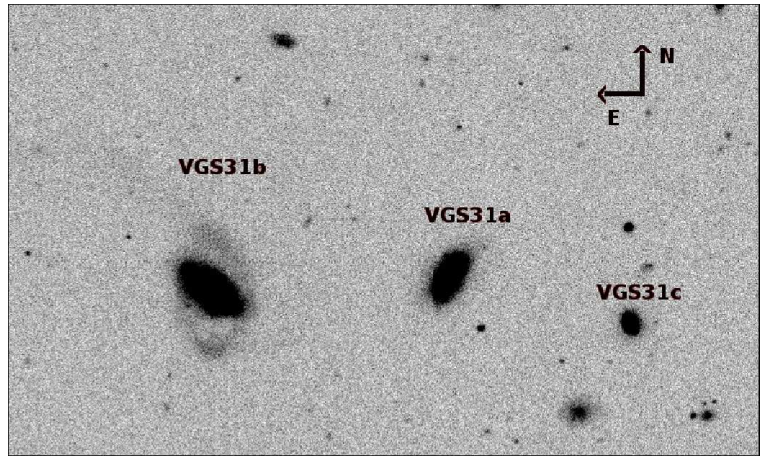}
\caption{SDSS g-band image of the galaxy group VGS31.} \label{VGS31}
\end{figure*}

VGS31 (Figure \ref{VGS31}) is a system of three aligned galaxies
belonging to a multiwavelength survey of 60 void galaxies, called
``The Void Galaxy Survey'' (VGS), conducted by \citet{Kreckel12}. 

\citet{Beygu13} show that the whole system is embedded in a common HI
envelope and the three galaxies have almost the same velocity. They
also assert that the absence of a velocity gradient throughout the HI
envelope suggests that it is a filament in which the three galaxies
are embedded. VGS31 consists of a central galaxy VGS31a and two
companions, VGS31b at the far east side and VGS31c at the far west
side, and shows strong signs of interaction. 

VGS31b is the most disturbed object of this small group; it is at a
distance of about 84 Mpc, based on $H_{0} = 75 \ km \ s^{-1}
\ Mpc^{-1}$ and has a heliocentric radial velocity of $V = 6218\ km
\ s^{-1}$, which implies that 1 arcsec = 0.4 kpc. VGS31b is a
starburst Markarian galaxy (MRK1477) with a one sided tail, curved
toward north-east, kinematically connected to the central
early-type disk galaxy (HG), a ring-like structure around the HG and a
fast rotating inner structure with streaming motions characteristic of
a bar. Both the tail and the ring are not detected in $H \alpha$ or
UV, but are clearly visible in optical and HI. Moreover,
\citet{Beygu13} also found that among the three galaxies, VGS31b has
higher $SFR_{\alpha}$.

The central HG have a diameter of $\sim$ 41 arcsec ($\sim 16$ kpc), while the
ring-like structure is more extended than the optical disk, having a
diameter of $\sim$ 60 arcsec ($\sim 24$ kpc).

\section{Observation and data reduction} \label{obs}

\subsection{Spectroscopic data}\label{spectra}

The long slit spectra analyzed in this work are archival data,
obtained with ISIS@WHT (Intermediate dispersion Spectrograph and
Imaging System) on May 2011 under the program ID. N20. ISIS is mounted
at the Cassegrain focus of the 4.2m William Herschel Telescope and is
a high-efficiency, double-armed, medium-resolution (8-120 \AA
$mm^{-1}$) spectrograph. Use of dichroic slides permits simultaneous
observing in both blue and red arms, which are optimised for their
respective wavelength ranges. The dataset studied in this work is the
same presented by \citet{Beygu13}, made by long-slit spectra acquired
along two different Position Angles (P.A.), $P.A. = 230^{\circ}$ and
$P.A. = 158^{\circ}$, corresponding to the central spheroidal galaxy
and the outer ring directions, respectively. These data were used by
\citet{Beygu13} to estimate the Star Formation Rate from the $H
\alpha$ luminosity. The main goal of the present study is to estimate
the metallicity in the ring-like structure of VGS31b, thus we have
analyzed only the spectra relative to this component, acquired with a
slit $1''$ wide and aligned at $P.A. = 158^{\circ}$.

At the systemic velocities of VGS31b, to cover the red-shifted
emission lines of $[OII]\lambda3727$, $H \gamma(\lambda4340)$,
$[OIII]\lambda4363$, $[OIII]\lambda\lambda4959,5007$, $H
\beta(\lambda4861)$ and $H \alpha(\lambda6563)$, the gratings R600B and
R1200R for the blue and the red arm were used, with a resolution of
0.45 and 0.26 \AA $pix^{-1}$ respectively.

The data reduction was carried out using the {\small CCDRED} package
in the IRAF\footnote{IRAF is distributed by the National Optical
  Astronomy Observatories, which is operated by the Associated
  Universities for Research in Astronomy, Inc. under cooperative
  agreement with the National Science Foundation.} ({\it Image
  Reduction and Analysis Facility}) environment. The main strategy
adopted for each dataset included dark subtraction\footnote{Bias
  frame is included in the dark frame.}, flat-fielding correction, sky
subtraction, and rejection of bad pixels. Wavelength calibration was
achieved by means of comparison spectra of CuNe+CuAr lamps acquired for
each observing night, using the IRAF TWODSPEC.LONGSLIT package. The
sky spectrum was extracted at the outer edges of the slit, for $r \ge
30$ arcsec from the galaxy center, where the surface brightness is
fainter than $24 mag/arcsec^2$, and subtracted off each row of the two
dimensional spectra by using the IRAF task BACKGROUND in the
TWODSPEC.LONGSLIT package.  On average, a sky subtraction better than
$1\%$ was achieved. The sky-subtracted frames were co-added to a final
median-averaged 2D spectrum.

\begin{figure*}
\centering
\includegraphics[width=15cm]{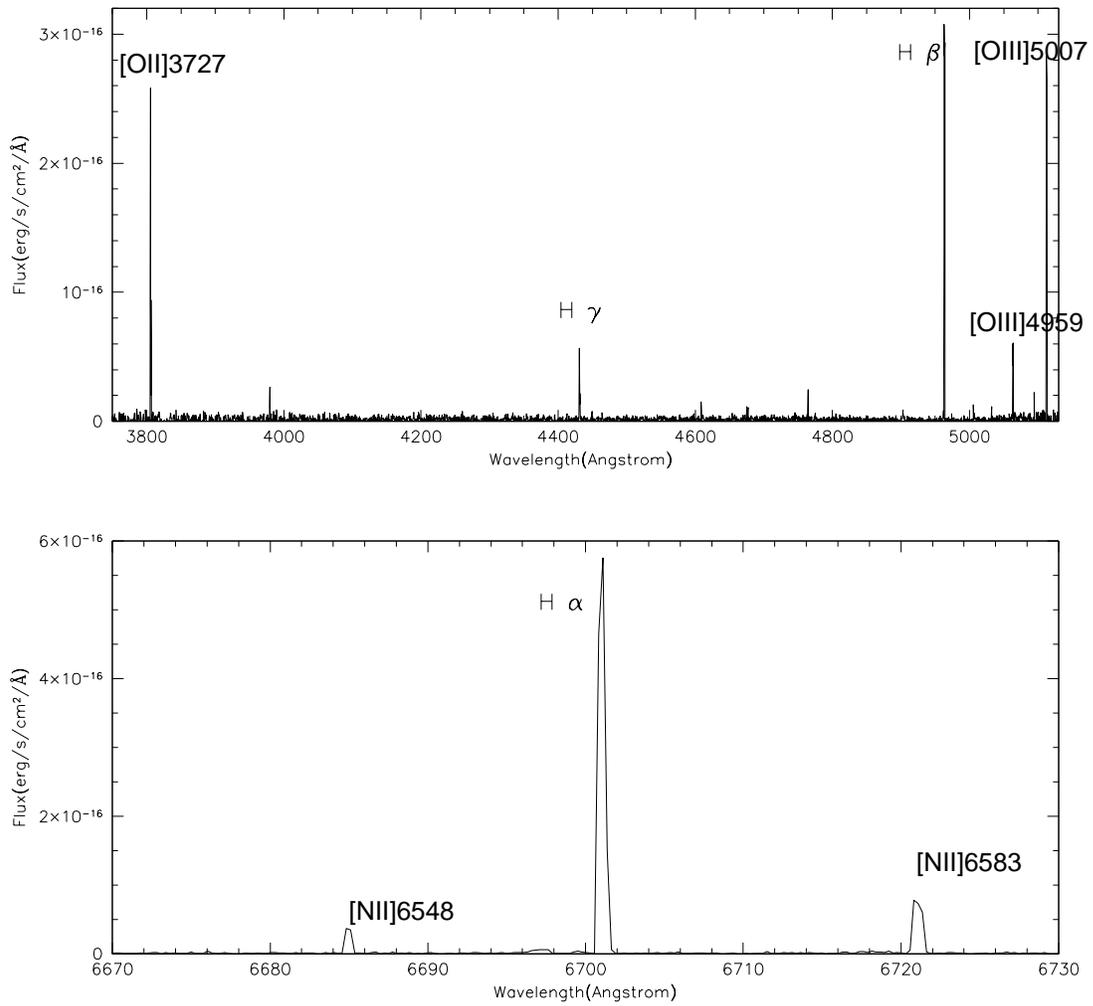}
\caption{1D spectra of VGS31b corresponding to a line extracted at a
  distance of $\sim$ 20 arcsec from the galaxy center.} \label{spec}
\end{figure*}

The final step of the data processing is the flux calibration of each
2D spectrum, by using observations of the standard star BD284211 and the
standard tasks in IRAF (STANDARD, SENSFUNC and CALIBRATE). To perform
the flux calibration, we extracted a 1-D spectrum of the standard star
to find the calibration function, and then we extracted a set of 1-D
spectra of the galaxy by summing up a number of lines corresponding to
the slit width. Since the slit width was $1 ''$ and the scale of the
instrument was $0.19 ''/pix$, we collapsed five lines to obtain each
1-D spectrum. Finally we applied the flux calibration to this
collection of spectra, showed in Fig.\ref{spec}.

The fluxes of these emission lines were measured using the IRAF
{\small SPLOT} routine, which provides an interactive facility to
display and analyze spectra. We evaluated the flux and equivalent
width by marking two continuum points around the line to be measured.
The linear continuum is subtracted and the flux is determined by
simply integrating the line intensity over the local fitted
continuum. The errors on these quantities are calculated, following
\cite{Per03}, by the relation $\sigma_{1} =
\sigma_{c}N^{1/2}[1+EW/(N\Delta)]^{1/2}$, where $\sigma_{1}$ is the
error in the line flux, $\sigma_{c}$ the standard deviation in a box
near the measured line and represents the error in the continuum
definition, N is the number of pixels used to measure the flux, EW the
equivalent width of the line, and $\Delta$ the wavelength dispersion
in \AA/pixel.

\subsubsection{Reddening correction} \label{red}
Reduced and flux-calibrated spectra and the measured emission line
intensities were corrected for the reddening, which account both for
that intrinsic to the source and to the Milky Way. By comparing the
intrinsic Balmer decrement $H \alpha/H \beta=2.89$, we derived the
visual extinction $A(V)$ and the color excess $E(B-V)$ by adopting
the mean extinction curve by \cite{Car89} $A(\lambda)/A(V) =
a(x)+b(x)R_{V}$, where $R_{V}[\equiv\ A(V)/E(B-V)]=3.1$ and
$x=1/\lambda$. All the emission lines in the spectra of VGS31b are in the
\emph{optical/NIR} range (see \citealt{Car89}), so we used the average
$R_{V}$-dependent extinction law derived for these intervals to
perform the reddening correction.

We derived the average observed Balmer decrements for the galaxy,
which are\\ $(H \alpha/H \beta)$ = $0.74
\pm\ 0.67$\\ $(H \gamma/H \beta)$ = $1.21
\pm\ 0.74$,\\ while the color excess obtained by using these observed
decrements are\\ $[E(B-V)]_{H \alpha}$ = $-1.34
\pm\ 0.39$\\ $[E(B-V)]_{H \gamma}$ = $-1.87 \pm\ 0.26$.\\ The
  negative value of the color excess indicates the
  presence of stars that are bluer and hotter than normal, and thus
  they have V apparent magnitude greater than the B one, leading to a
  negative B-V color. In general, stars hotter than Vega, which have
  $E(B-V)=0$, have negative color excess \citep{Mas98}.

Such values of E(B-V) are used to derive the extinction $A_{\lambda}$
through the Cardelli's law. Finally, the corrected fluxes are given by
the following equation and listed in Tab.\ref{fluxes}.

\begin{equation}
\frac{F^{\lambda}_{int}} { F^{H \beta}_{int}} = \frac{F^{\lambda}_{obs}}  {F^{H \beta}_{obs}} 10^{0.4[A_{\lambda}-A_{H \beta}]}.
\end{equation}

\subsection{Photometric data}

We used SDSS images of VGS31 observed with a 2.5 meters telescope in
the {\it u, g, r, i, z} bands. The field of view was 3 square degrees,
with a pixel scale of 0.396 arcsec/pixel and an exposure time of about
53 seconds.

\begin{figure*}
\centering
\includegraphics[width=8cm]{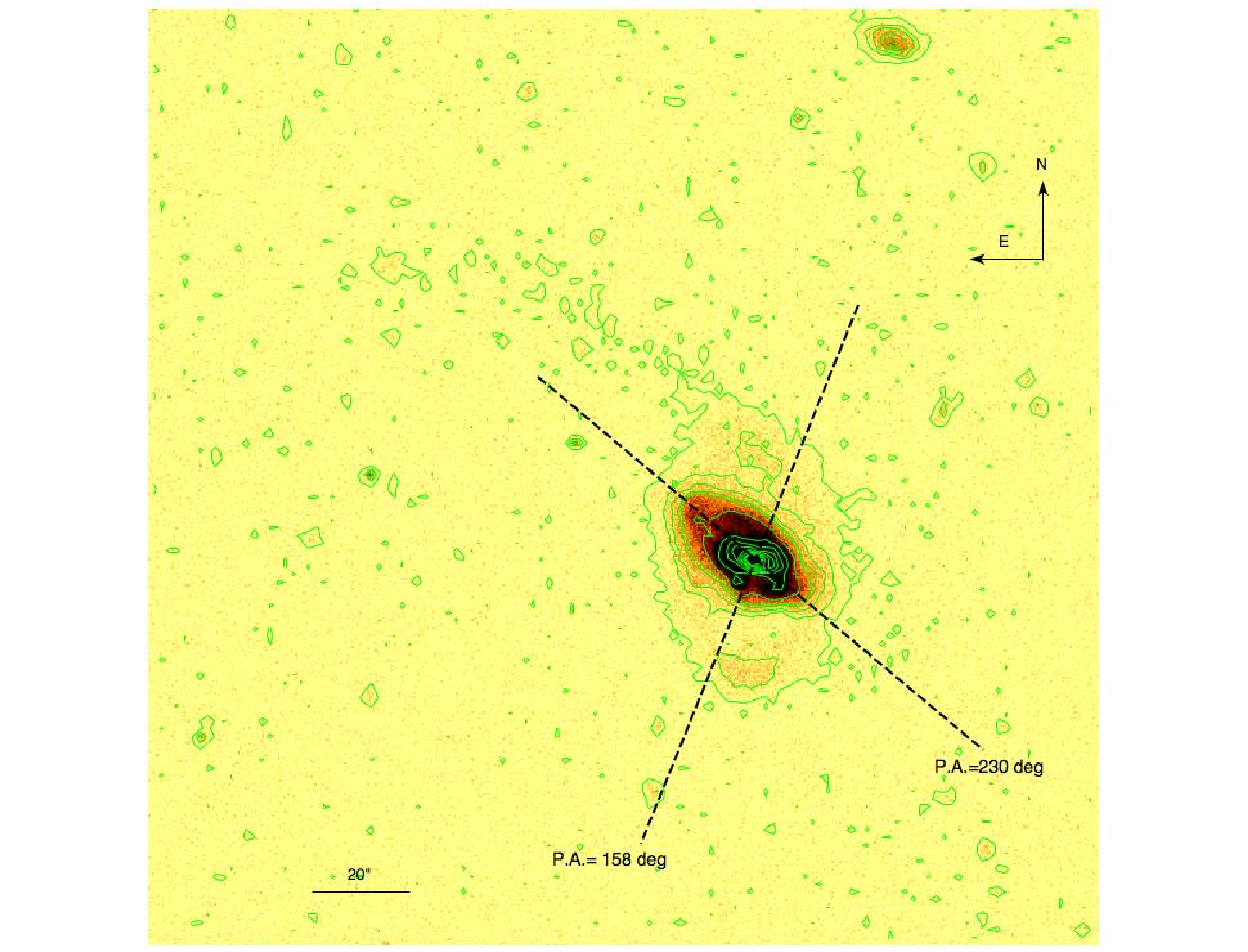}
\includegraphics[width=8cm]{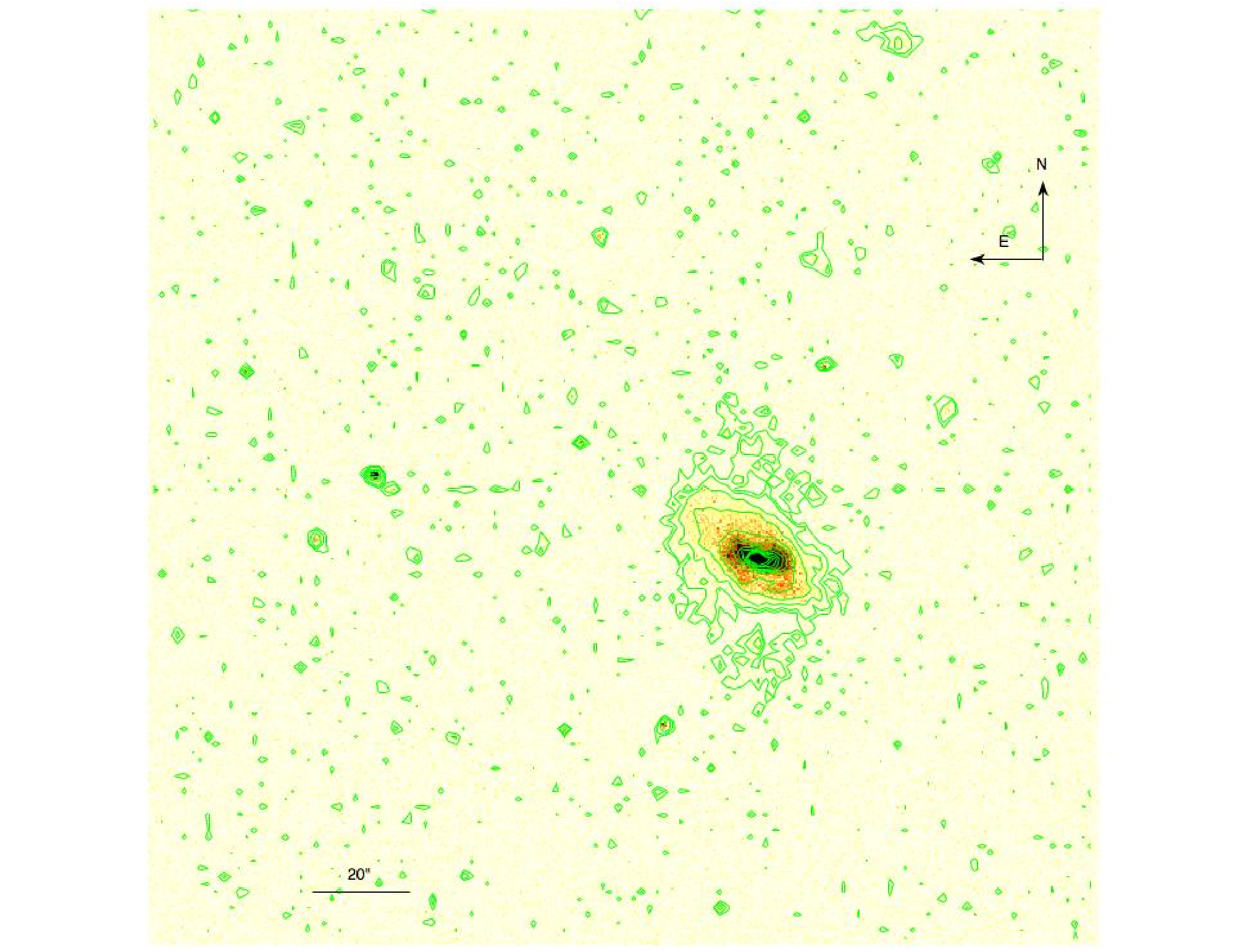}
\caption{g band (left) and z band (right) images of VGS31b with,
  superimposed, the contour levels. The dashed lines correspond
    to the directions where the archive spectra are taken and where the
    light profiles have been extracted (see Sec.\ref{mag}).}
  \label{VGS31b}
\end{figure*}

The ring-like structure of the galaxy is clearly visible in
the {\it g, r} and {\it i} bands, while it disappears in the {\it u}
and {\it z} bands (see Fig. \ref{VGS31b}). 

\section{Host galaxy and ring morphology}\label{morph}

The SDSS images of VGS31b (see Fig. \ref{VGS31b}) show that most of
the galaxy light comes from the central HG, which has the morphology
of an early-type disk galaxy and an average diameter of $\sim$ 16
kpc. The ring-like structure, visible along the North-South direction,
with a diameter of about 24 kpc, is more extended in radius than the
host galaxy. The ring is clearly visible in the {\it g, r} and {\it i}
bands, while it is not detectable in the {\it u} and {\it z} bands,
and it appears knotty and dusty.

\begin{figure*}
\centering
\includegraphics[width=10cm, angle=90]{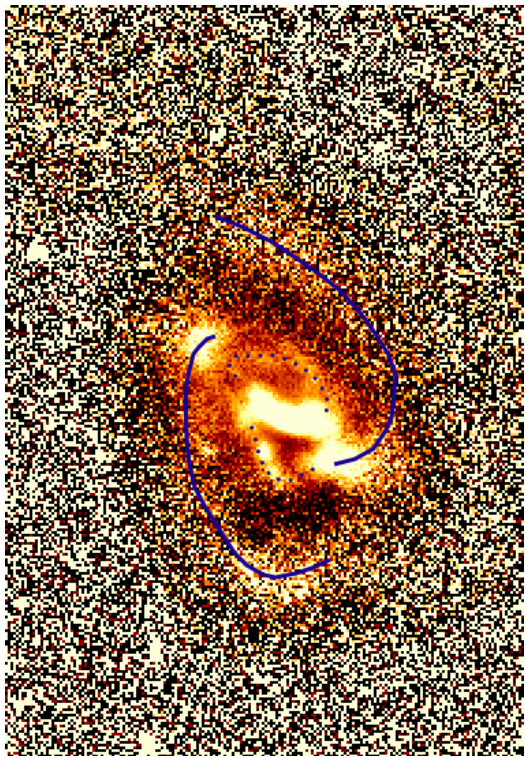}
\caption{High frequency residual g band image for VGS31b. Lighter
  colors correspond to brighter features. The blue dotted and solid
  curves, highlight the two arcs connecting the edges of the central
  bar and the two arms of the ring, respectively. The north is up while
  the east is on the left.} \label{ratio}
\end{figure*}

To emphasize the galaxy substructure, we create a residual image
produced by taking the ratio of the original reduced g-band image with
a smoothed one.  We use the IRAF task {\small FMEDIAN} to smooth the
original reduced image, by using a window size of $11 \times 11$, that
is chosen to best emphasize the inner structure of the central
host. The final un-sharp masked image of VGS31b is shown in
Fig.\ref{ratio}: here, the complex structure of this galaxy, even in
the very central regions, stands out very clear. We can identify two
main features: {\it i)} a luminous bar-like structure which crosses the
center of the galaxy, along NE-SW direction, $P.A. \sim 230^{\circ}$
(see also Sec.\ref{phot}), and {\it ii)} two very luminous blobs
observed in the NE and SW sides of the bar. Such features are also
detected in the g-band light profiles (see Figure
\ref{prof_g}). Moreover, there are other two fainter substructures,
almost parallel to the central luminous bar-like features, which
appear as two arcs that connect the edges of the bar; that on the NE
side appear more knotty, suggesting the presence of possible
star-forming regions. On the other hand, the outer ring-like structure
appears very smooth. The two bright blobs seem to be the
``starting-point'' of the two arms of the ring, since they are located
on the same apparent ellipse (see Fig.\ref{ratio}). The two inner arcs
seems to connect the above features to the galaxy center by tracing a
``spiral-like'' pattern.

\section{Photometry: Light and color distribution}\label{phot}

A quantitative photometry of VGS31b has been performed by using all
the SDSS bands (see Sec.\ref{obs}) in order derive the total light and
typical scale-lengths of the main components observed in VGS31b (host
galaxy and ring), to locate the inner substructure and, where
possible, to estimate their contribution to the total light. To this
aim, in the following sections we describe the isophotal analysis and
light distribution.

\subsection{Isophotal analysis} 

We used the {\small IRAF-ELLIPSE} task on the SDSS images to perform
the isophotal analysis for VGS31b and the results are shown in Figure
\ref{phot}. The average surface brightness extends up to about 31
arcsec from the galaxy center for the {\it g, r} and {\it i} bands,
while the {\it u} and {\it z} bands are less extended, reaching 18 and 28
arcsec respectively; the half-light radius is $R_{e}=8.3$ arcsec in
the {\it u} and {\it g} bands, and $R_{e}=9.1$ in the {\it r, i} and
{\it z} ones.

\begin{figure*}
\centering
\includegraphics[width=12cm]{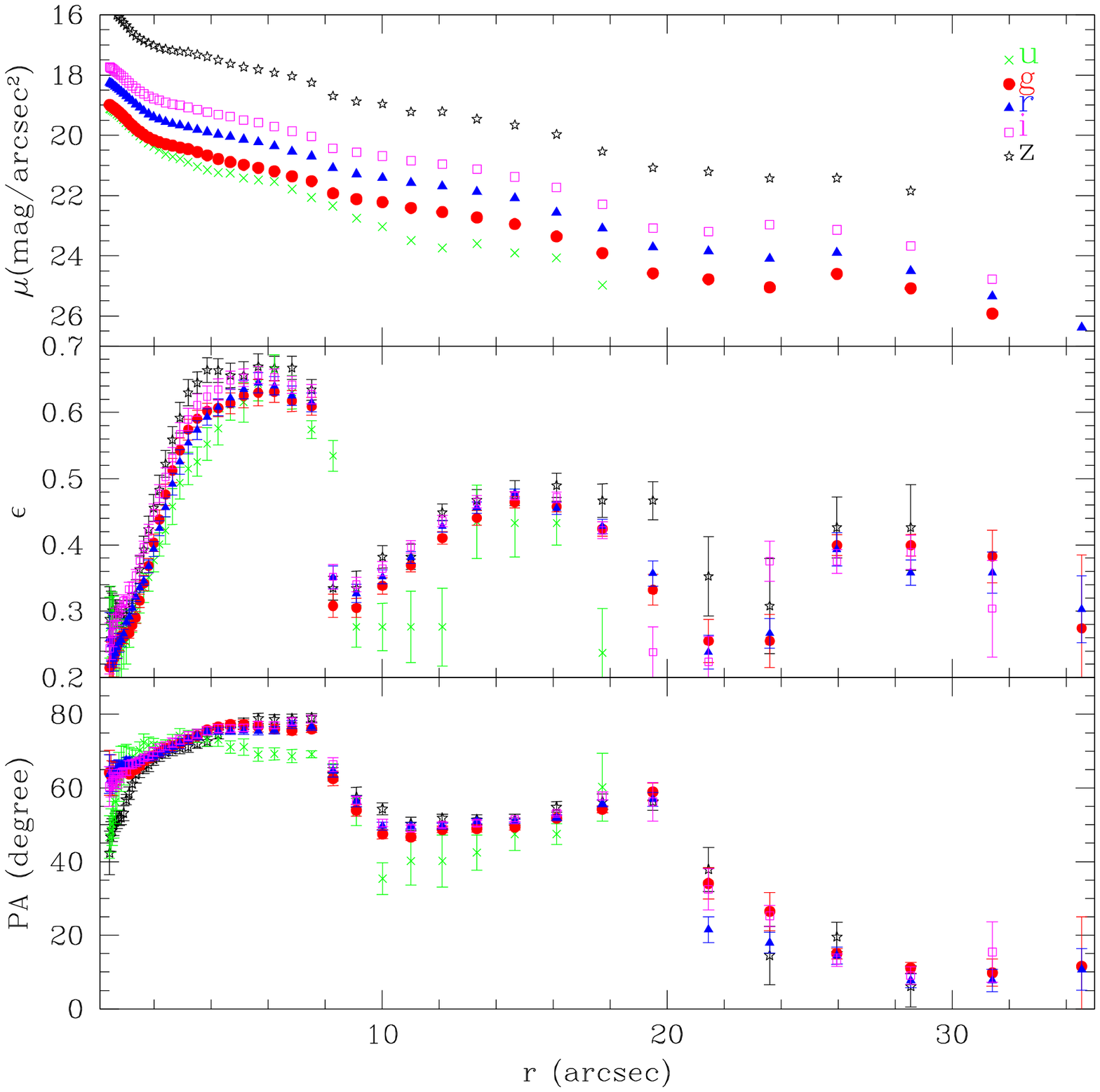}
\caption{Position Angle (P.A.), Ellipticity ($\epsilon$) and mean
  surface brightness profile in the u, g, r, i and z SDSS bands. The
  error bar for the surface brightness profile ($\pm$ 0.04 for the u
  band, $\pm$ 0.01 for the g, r and i bands, and $\pm$ 0.02 for the z
  band) is within the dimensions of data points.} \label{phot}
\end{figure*}

For a semi-major axis r, in the range $0\leq r \leq 8$ arcsec, the
Position Angle (P.A.) is almost constant and equal to $\sim
70^{\circ}$, indicating that in this regions the isophotes are almost
coaxial; the ellipticity rises from 0.2 to 0.7, showing the presence
of a flatter structure around 8 arcsec from the galaxy center. For
$8\leq r \leq 20$ arcsec the profiles reflect the presence of the ring
and we observe a twisting of the isophotes of about $30^{\circ}$ and
lower values of the ellipticity ranging from 0.3 to 0.5. Finally, for
$r \geq 20$ arcsec we observe another change in both ellipticity and
P.A. with $\epsilon \sim$ 0.4 and a twisting of $\sim
50^{\circ}$. At this radii the ellipticity and the P.A. are those
  of the isophotes corresponding to the ring-like structure, while in
  the range $0\leq r \leq 8$ arcsec they correspond to the host
  galaxy. The difference between the P.A. of the central galaxy and of
  the ring-like structure is $\sim 70^{\circ}$, which is the relative
  inclination between host galaxy and ring. This difference turns to
  be consistent with that between the two P.A.s used used to acquire
  the spectra along the two components, i.e. HG and ring, (see
  Sec. \ref{spectra}).

\begin{figure*}
\centering
\includegraphics[width=12cm]{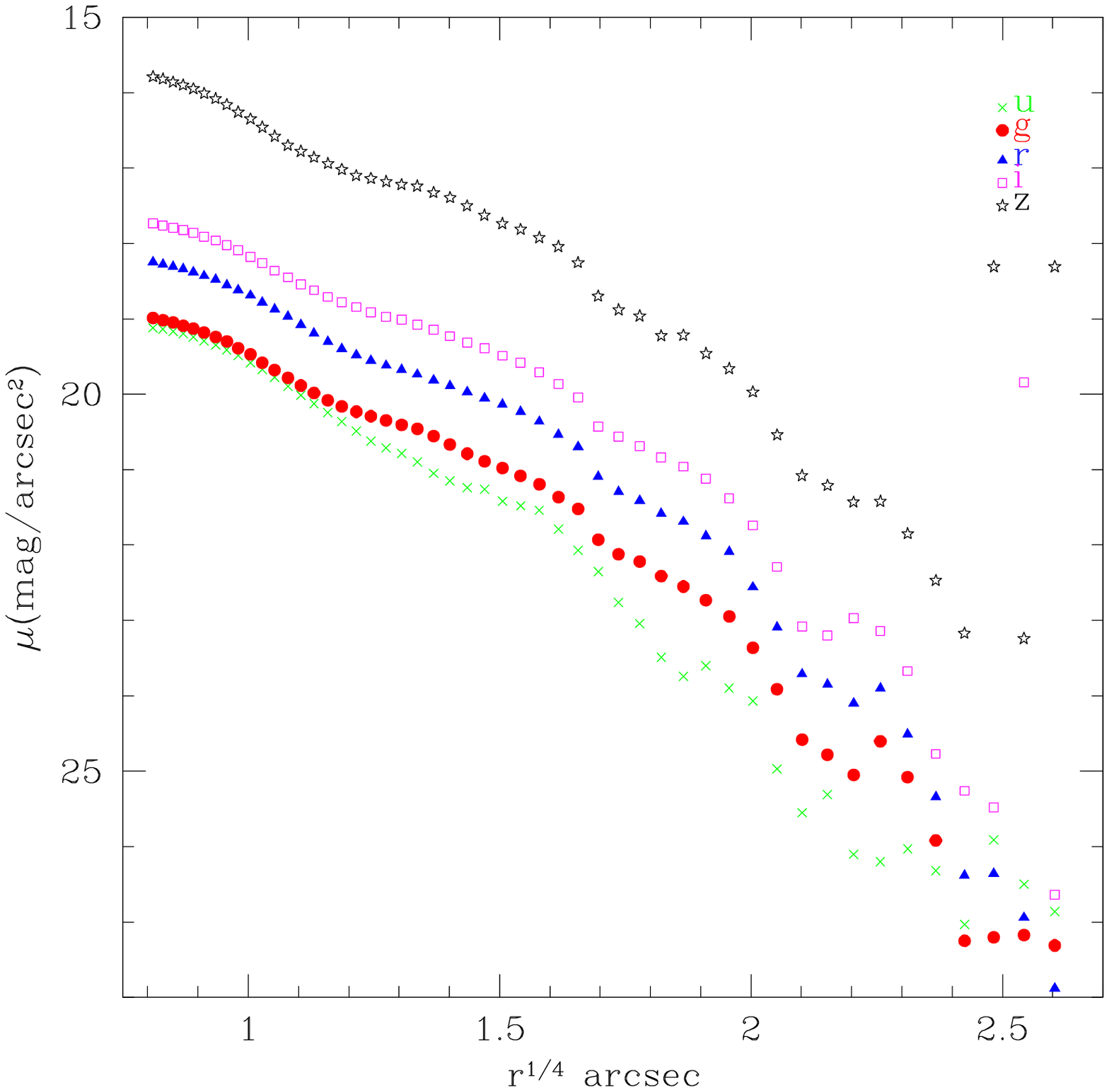}
\caption{de Vaucouleur radial surface brightness profile. The error
  bar ($\pm$ 0.04 for the u band, $\pm$ 0.01 for the g, r and i bands,
  and $\pm$ 0.02 for the z band) is within the dimensions of data
  points.} \label{r1/4}
\end{figure*}

The ring-like structure affects the surface brightness profiles, in
all bands, at least at three different distance from the galaxy
center: at $r \sim 5$ arcsec, $r \sim 10.5$ arcsec, $r \sim 28$
arcsec, see Figure \ref{r1/4}. These features turns to be much more
evident in the light profiles extracted along the major axes of the
two components (see Figures \ref{prof_g} and \ref{prof_i}).

\begin{figure*}
\centering
\includegraphics[width=12cm]{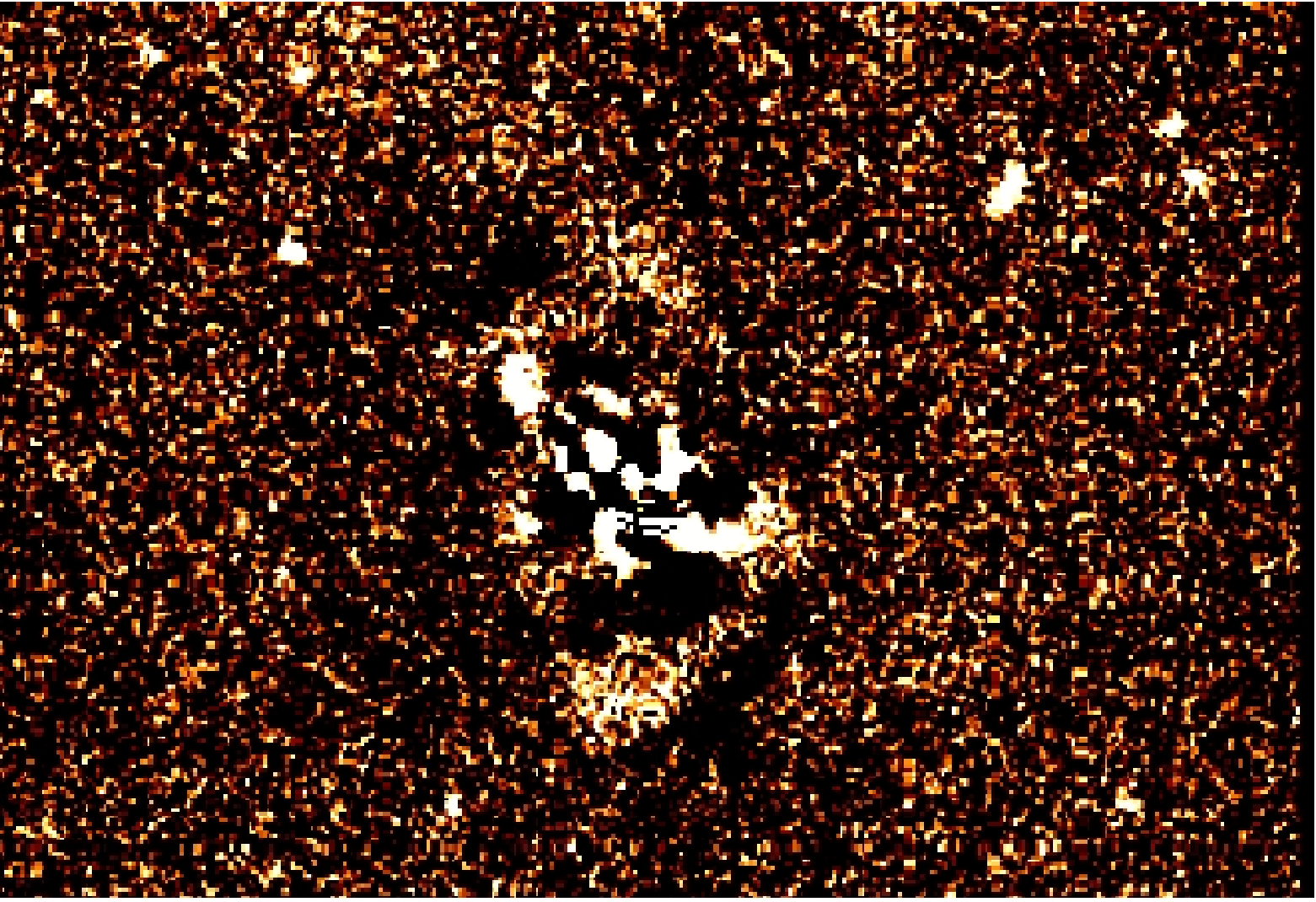}
\caption{Residual of the subtraction of the model to the g band
  image.} \label{res}
\end{figure*}

In order to analyze the complex structure of VGS31b, which is very
luminous in the g band, we have also derived the 2D model in this band
by using the IRAF task {\small BMODEL}. We have created a
2-dimensional model image from the results of the isophotal analysis
generated by the isophote fitting task {\small ELLIPSE} and subtracted
this model to the g band image, obtaining the residual image shown in
Fig. \ref{res}. The residuals emphasize the presence of a very complex
structure in this galaxy, as already shown by the un-sharp masked
image. In particular, also in this case we can identify the ring's
arms connected to the two bright blobs at the ends of the bar-like
structure, and the ``spiral-like'' pattern traced by the ring up to
the central regions of the galaxy.

\subsection{Light and color distribution}\label{mag}

\begin{figure*}
\centering
\includegraphics[width=10cm]{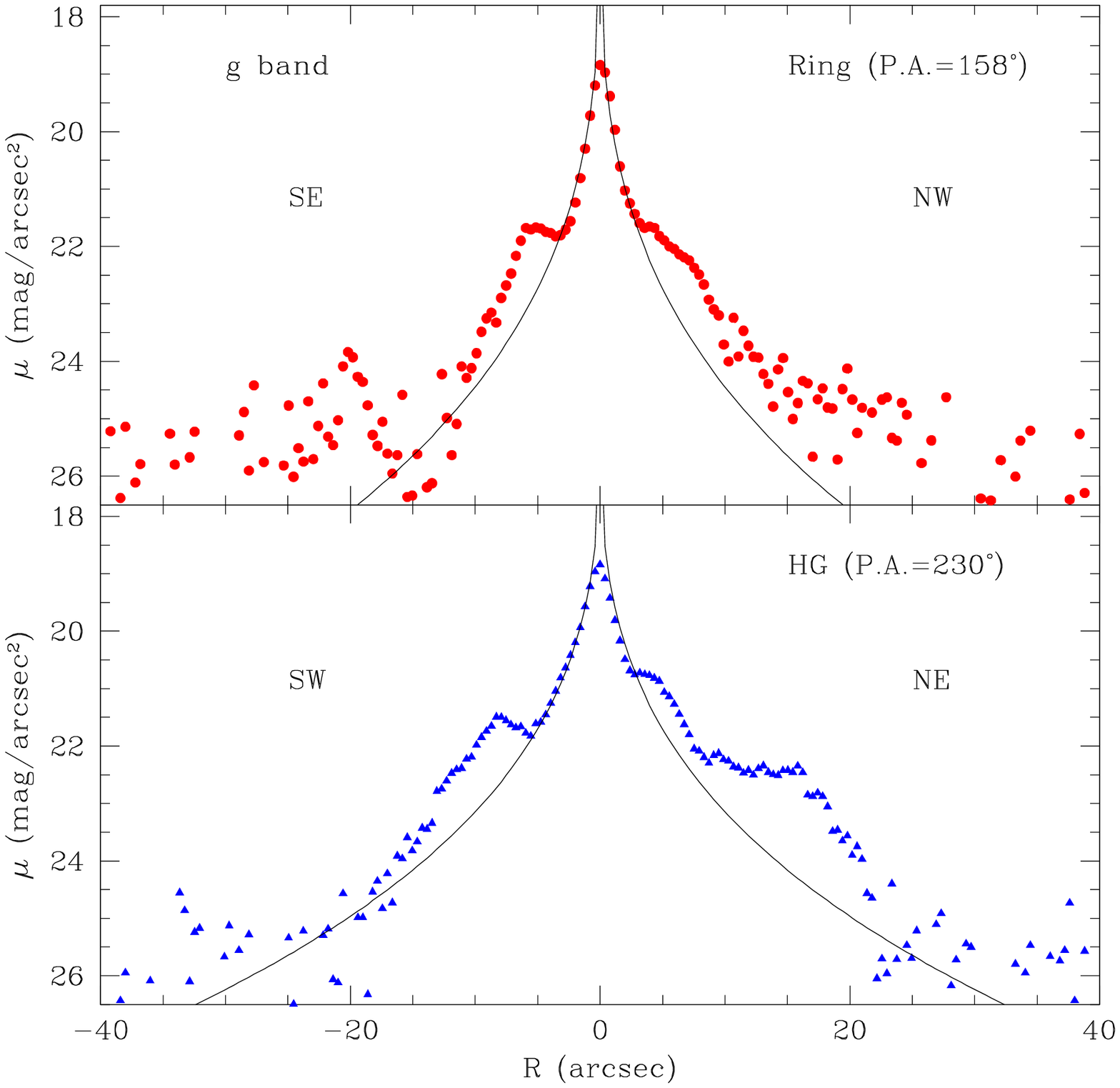}
\caption{{\it Top panel} - g-band light profile along the ring-like
  component (P.A. 158 deg). {\it Bottom panel} - g-band light profile
  along the host galaxy (P.A. 230 deg). The black line is the 1D model
  of the light profile. The error bar on the surface brightness ($\pm$
  0.01) is within the dimensions of data points.} \label{prof_g}
\end{figure*}

\begin{figure*}
\centering
\includegraphics[width=10cm]{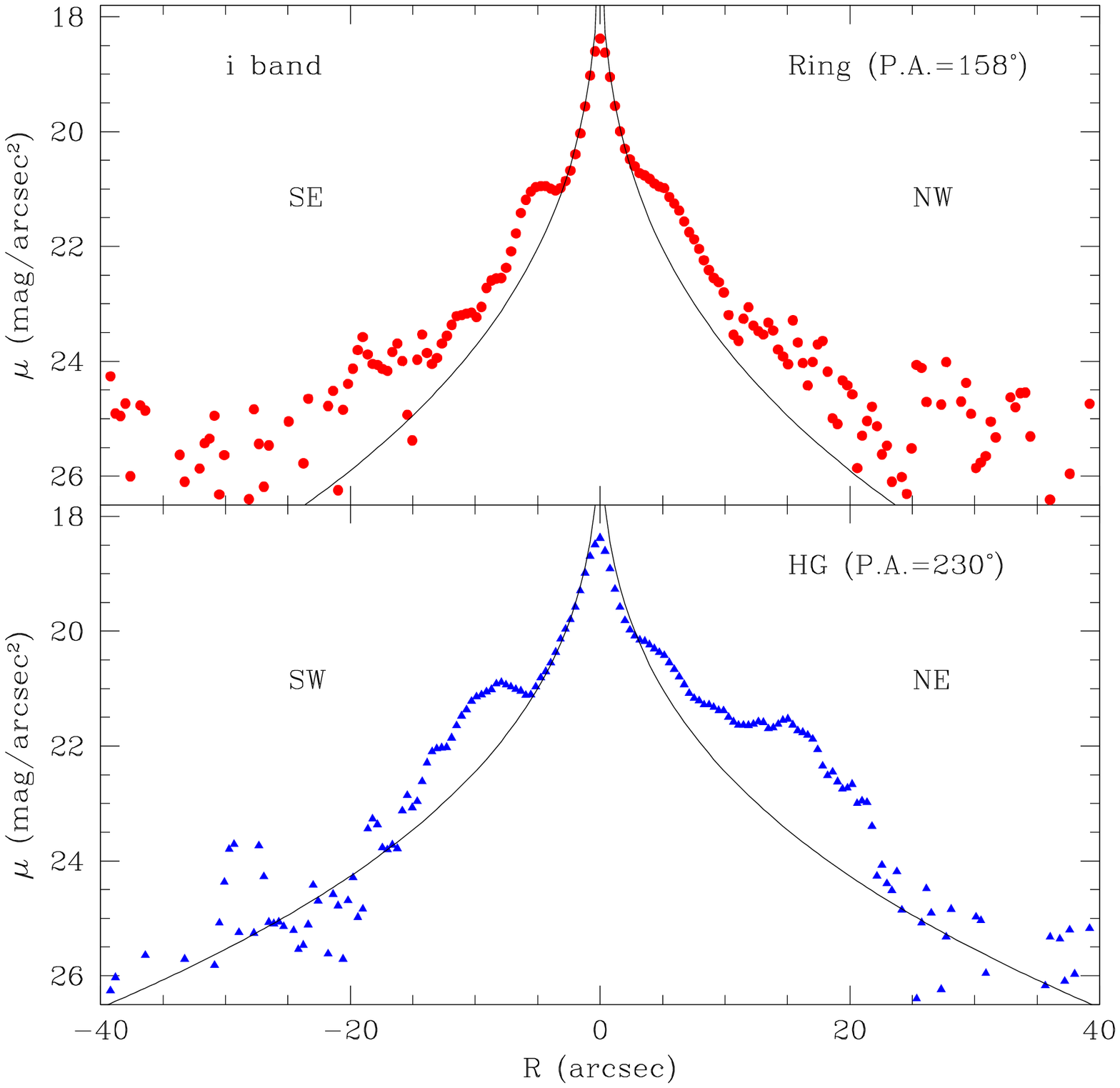}
\caption{{\it Top panel} - i-band light profile along the ring-like
  component (P.A. 158 deg). {\it Bottom panel} - i-band light profile
  along the host galaxy (P.A. 230 deg). The black line is the 1D model
  of the light profile. The error bar on the surface brightness ($\pm$
  0.01) is within the dimensions of data points.} \label{prof_i}
\end{figure*}

In Figures \ref{prof_g} and \ref{prof_i} we show the light profiles in
the g and i bands, respectively, along the HG and ring-like structure,
$P.A. = 230$ and $P.A. =158$ degrees respectively, at the same
position angles used to acquire the spectra analyzed in this work (see
Sec. \ref{spectra} and Fig.\ref{VGS31b}). From this point on, we refer
as ``major axis'' of each component in VGS31b, i.e. HG and ring, the
directions along the above two position angles. In both bands, along
the HG major axis, the light profiles show a central bright
concentration inside 4 arcsec, the three bright bumps, already
identified in the average isophotal profiles (see Fig.\ref{r1/4}),
located at $4 \le R \le 5$ arcsec and $10 \le R \le 18$ arcsec on the
NE side, and at $R \sim 10$ arcsec on the SW side. The inner one,
reflects the presence of the inner bar-like structure, while the other
two ones, at a larger radii, are due to the bright blobs identified in
Fig.\ref{ratio}, which are probably connected to the ring.  Along the
ring major axis, light profiles are smoother, showing a behaviour
typical for galaxies with rings \citep{Iod02a}: two prominent bumps,
at $4 \le R \le 8$ arcsec, are due to the ring light superimposed to
the underlying HG light. Both the light distribution along the
  main axes of the galaxy and the previous analysis of the galaxy
  morphology (see Sec.\ref{phot}) show that the ring contributes to
  the galaxy light at several distances from the centre. Since we aim
  to derive an average estimate of the metallicity (Z) for the ring
  component, we modelled the underlying light distribution of the HG,
  in order to obtain the range of radii where the ring dominates with
  respect to the HG and derive the average value for Z.  Such model of
  the light distribution of the only HG component need to be performed
  by fitting the 1D light profiles, where is possible and easy to
  exclude the regions where the ring dominates, i.e.  at $4 \le R \le
  5$ arcsec and $10 \le R \le 18$ arcsec on the NE side, and at $R
  \sim 10$ arcsec on the SW side.  The light distribution of the HG in
  VGS31b has been modelled by adopting a Sersic law $\mu [r] \propto
  \mu_e + (r/r_e)^{1/n}$ \citep{Sersic63}, both in the g and in the i
  bands: results are shown in Figures \ref{prof_g} and
  \ref{prof_i}. The best fit is obtained with $\mu_e = 22.3 \pm 0.1$,
  $r_e=6.8 \pm 0.5$, $n=3.0 \pm 0.1$, and suggests that: {\it i)} both
  along the HG and ring major axis (bottom and top panels,
  respectively), in the central regions, for $R \le 5$ arcsec, the HG
  is the dominant component; {\it ii)} along the HG major axis (bottom
  panels), in the range $0.5 \le R \le 20$ arcsec, most of the light
  comes from the ring, while at larger distances, inside the scatter
  of the data, the light distribution follows the fitted Sersic law
  for the HG; {\it iii)} along $P.A.=158^\circ$ (top panels), the ring
  is the dominant component in the light distribution at all distance
  for $R \ge 5$ arcsec.

\begin{figure*}
%\centering
\includegraphics[width=8cm]{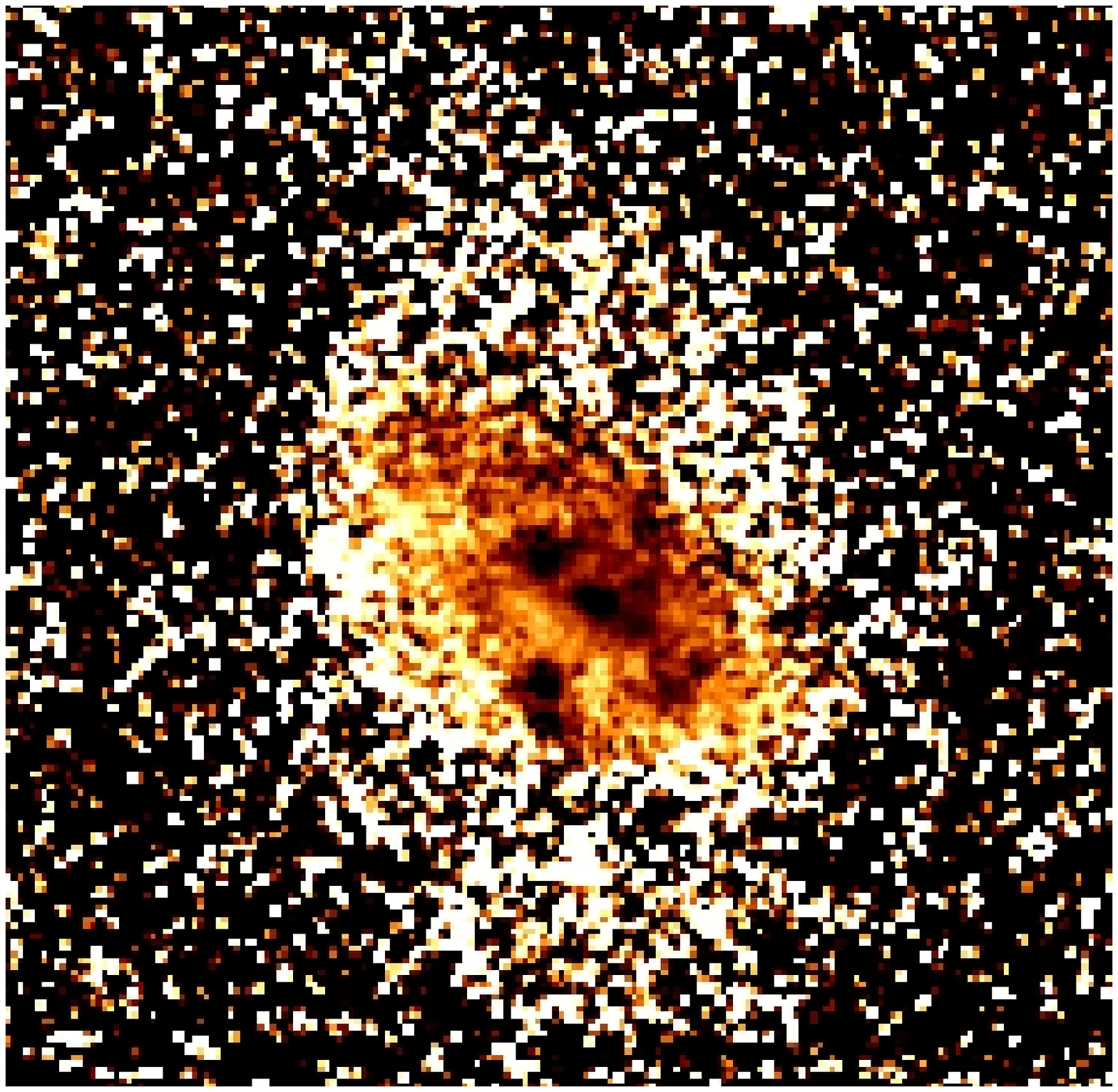}
\includegraphics[width=8.2cm]{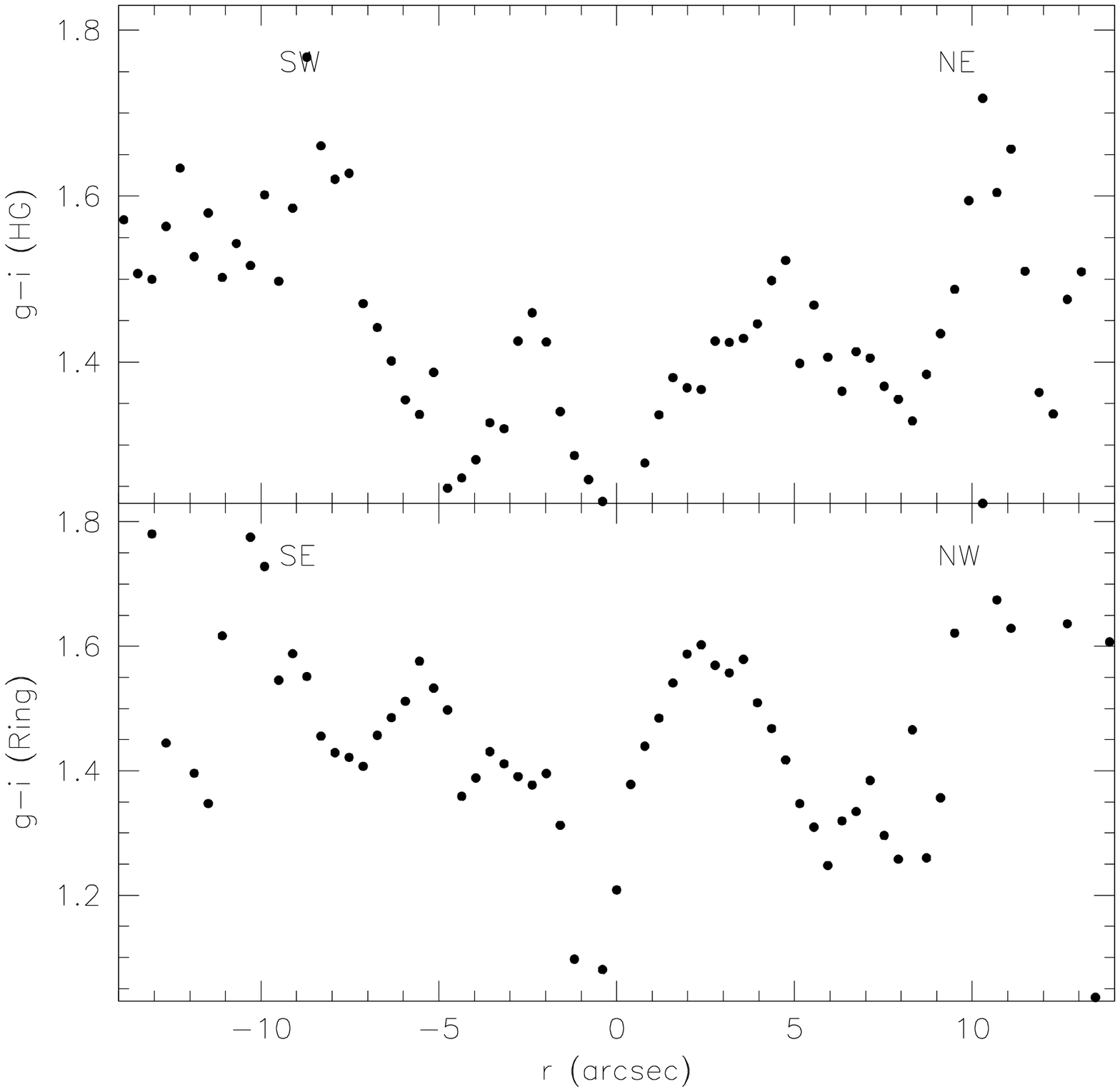}
\caption{{\it Left panel} - g-i color map. The North is up, while the
  East is on the left of the image. Lighter colors correspond to
  redder galaxy regions. {\it Right panel} - g-i color profile along
  the major axis of the central disk galaxy (top) and along the major
  axis of the ring-like structure (bottom). The error bar ($\pm$ 0.02)
  is within the dimensions of data points.} \label{gi}
\end{figure*}

\begin{figure*}
\centering
\includegraphics[width=8cm]{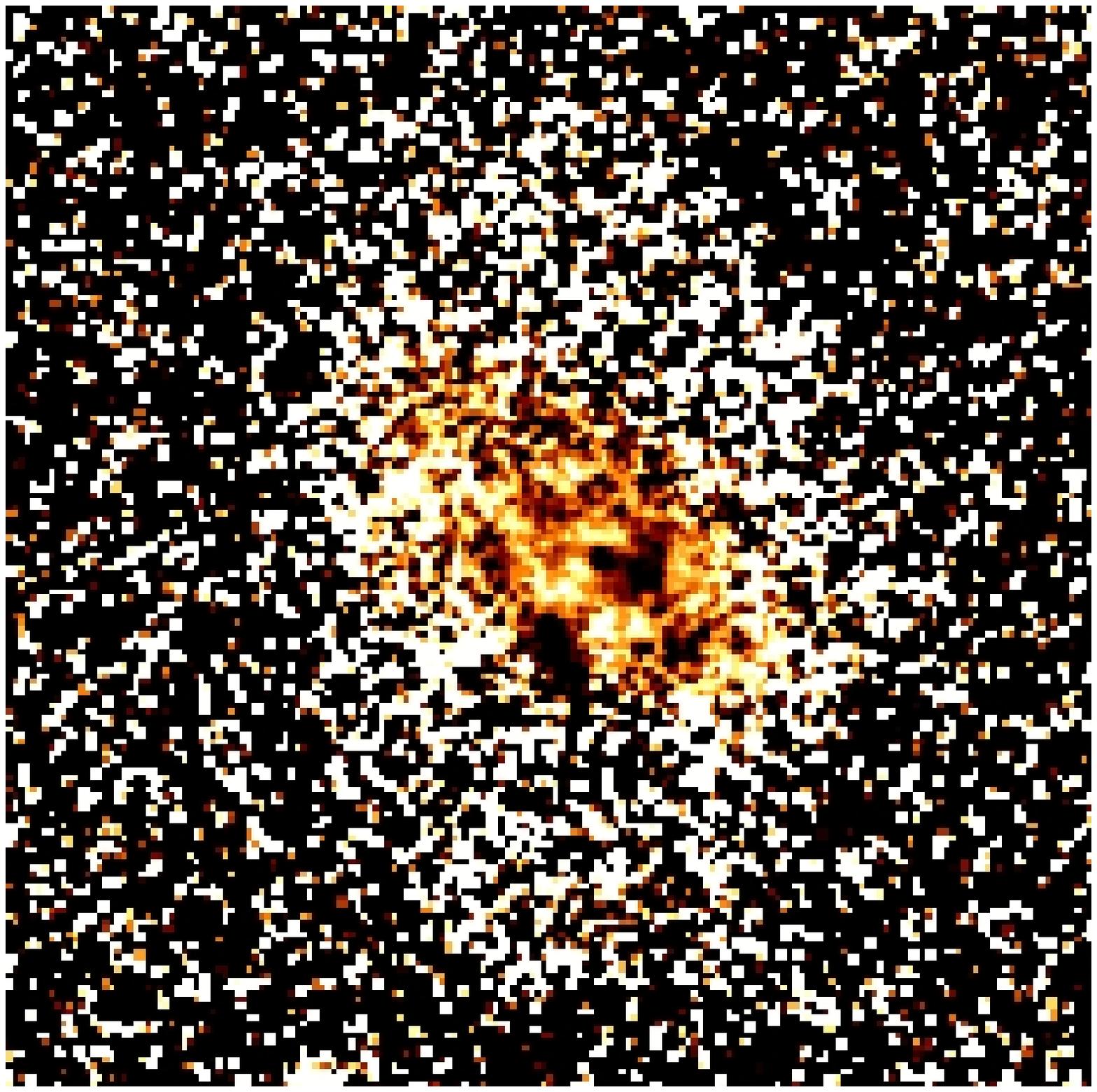}
\includegraphics[width=8.2cm]{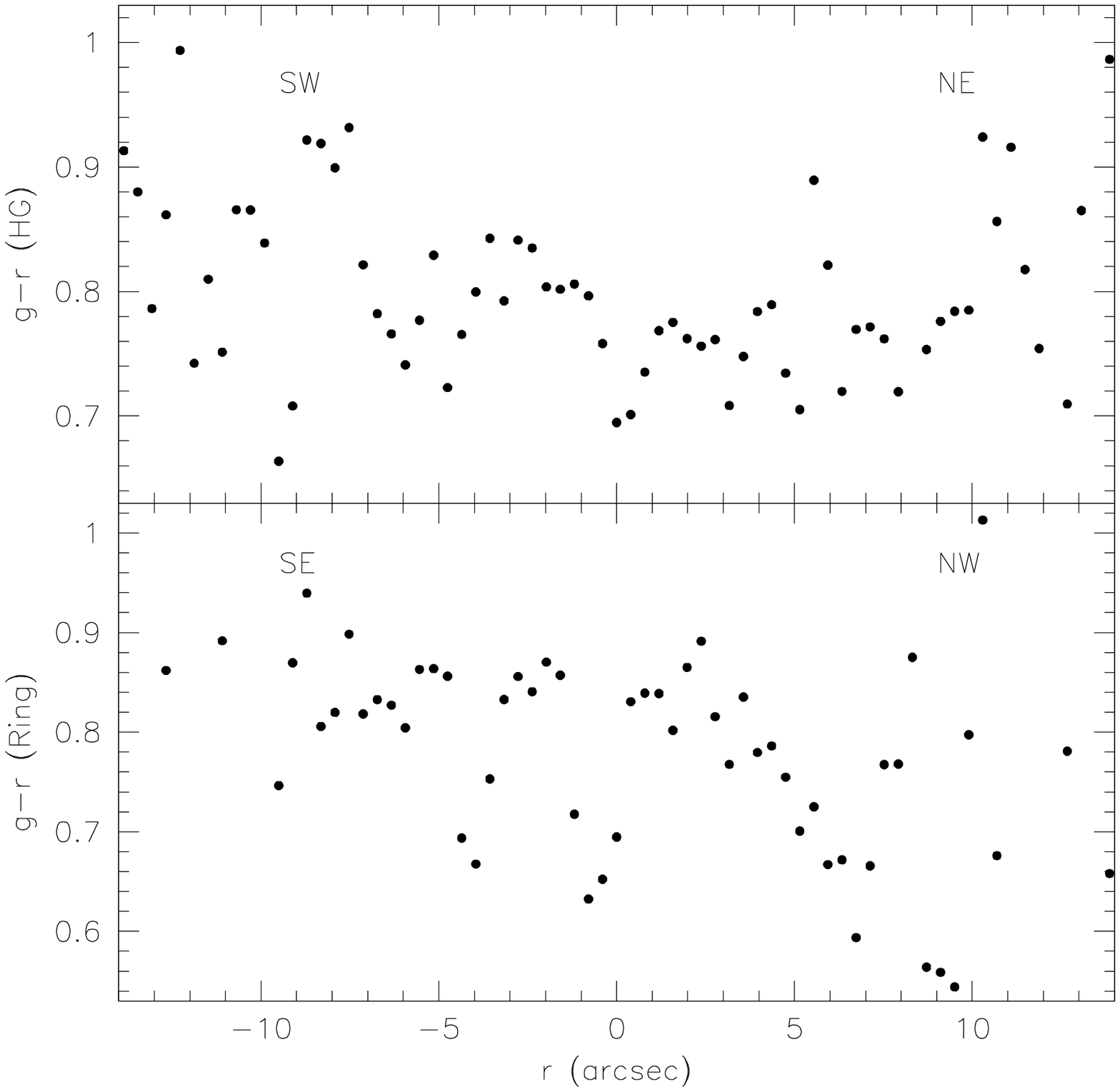}
\caption{{\it Left panel} - g-r color map. The North is up, while the
  East is on the left of the image. Lighter colors correspond to
  redder galaxy regions. {\it Right panel} - g-r color profile along
  the major axis of the central disk galaxy (top) and along the major
  axis of the ring-like structure (bottom). The error bar ($\pm$ 0.02)
  is within the dimensions of data points.} \label{gr}
\end{figure*}

We have also derived the g-i and g-r color profiles along the host
galaxy and ring major axis (Figures
\ref{gi} and \ref{gr}) and the 2-dimensional color maps. 
On average, the central regions of the galaxy
have bluer colors, with a maximum value of g-i $\sim 1.6 \pm 0.02$ for
the host galaxy and $\sim 1 \pm 0.02$ for the ring, and g-r $\sim 0.7
\pm 0.02$ for the host galaxy and $\sim 0.6 \pm 0.02$ for the
ring. Both color maps are characterized by several areas of very
blue colors (the nucleus, inside 4 arcsec from the center, two knots
along the HG major axis and one more on the East side with respect to
the galaxy nucleus), which suggest the presence of star forming
regions, and this turns to be consistent with the results previously
found by \citet{Beygu13}.

We also derived the integrated magnitudes and {\it g-r} and {\it g-i}
colors in three polygons including the tail at the North-East side of
VGS31b and the two sides of the ring. The polygons are determined from
the g band image, using the IRAF task {\small POLYMARK}, and used for
all bands. The integrated magnitudes inside each polygon are evaluated
using the IRAF task {\small POLYPHOT}. The derived magnitudes and
colors are reported in Table \ref{magnitudes}.

\begin{table*}
\caption{\label{magnitudes}Integrated magnitudes and colors of different regions of VGS31b.} 
\centering
\begin{tabular}{lccccccc}
\hline\hline
Component& Region & $m_{g} (mag)$ & $m_{r} (mag)$ & $m_{i} (mag)$ &  g-r & g-i\\
         &        & $\pm 0.01$   & $\pm 0.01$   & $\pm 0.01$ & $\pm 0.02$& $\pm 0.02$\\
\hline
Tail & NE & 18.12&17.57 & 16.32 & 0.55 & 1.8\\
Ring & N & 18.12 &17.20 & 16.19 & 0.92 & 1.9\\
Ring & S & 18.10 &17.53 & 16.60 & 0.57 & 1.5\\
\hline
\end{tabular}
\end{table*}

The stellar population synthesis model by \citet{Bru03} were used to
reproduce the integrated colors in the selected regions, in order to
estimate the average stellar mass in the tail and in the
ring-like structure of VGS31b. The key input parameters for GISSEL
({\it Galaxies Isochrone Synthesis Spectral Evolution Library},
\citealt{Bru03}) are the Initial Mass Function (IMF), the Star
Formation Rate (SFR) and the metallicity. For the ring-like structure
and the tail of VGS31b, since they have bluer colors than the host
galaxy, which suggest a younger age for this component, we used
models with linearly declining SFR ($\psi(t) =
2M_{\star}\tau^{-1}[1-(t/\tau)]$) computed for metallicities of
Z=0.008 and Z=0.0004, because these models reproduce the integrated
colors of late-type galaxies. In every model it has been assumed
that stars form according to the \citet{Sal55} IMF, in the range from
0.1 to 125 $M_{\odot}$.

For the tail we derive a stellar mass of the order of $10^{8}
M_{\odot}$, while for the ring the estimated mass is of about $4
\times 10^{8} M_{\odot}$. The implications of these results will be
discussed in section \ref{conc}.

\section{Empirical oxygen abundances determination}\label{oxy}

In order to derive the oxygen abundance $12+log(O/H)$ and the
metallicity of the ring-like structure in VGS31b, we have measured the
\emph{Oxygen abundance parameter} $R_{23} = ([OII]\lambda 3727 +
     [OIII]\lambda \lambda 4959 + 5007)/H \beta$ \citep{Pag79}, by
     following the procedure outlined by \cite{Spav10,Spav11}, adopted
     for other PRGs galaxies.

The \emph{Empirical methods} are based on the cooling properties of
ionized nebulae, which translate into a relation between emission-line
intensities and oxygen abundance. Several abundance calibrators have
been proposed based on different emission-line ratios: $R_{23}$
\citep{Pag79} and $S_{23}$ \citep{Diaz00}. Among the others, in this work
we used the so-called P-method introduced by \cite{Pil01}.

\cite{Pil01} realized that, for fixed oxygen abundances, the value
of $X_{23} = log R_{23}$ varies with the excitation parameter $P =
R_{3}/R_{23}$, where $R_{3} = OIII[4959+5007]/H \beta$, and proposes
that this latter parameter could be used in the oxygen abundance
determination. This method, called the ``P-method'', proposes using a more
general relation of the type $O/H = f(P, R_{23})$, than the
relation $O/H = f(R_{23})$ used in the $R_{23}$ method. The equation
related to this method is
\begin{equation}\label{pil-cal}
    12+log(O/H)_{P} = \frac{R_{23}+54.2+59.45P+7.31P^{2}}{6.07+6.71P+0.371P^{2}+0.243R_{23}}
\end{equation} where $P = R_{3}/R_{23}$. It can be used for oxygen abundance
determination in moderately high-metallicity HII regions with
undetectable or weak temperature-sensitive line ratios
\cite{Pil01}. The definition of moderately high metallicity is adopted
from \citet{Pil01} and refers to the abundance interval $8 < 12+logO/H
< 8.5$ where the relation $O/H = F(R_{23})$ \citep{Pag79}
degenerates. As suggested by \citet{Pil01}, the positions of HII
regions in the $P-R_{3}$ diagram are related to their oxygen
abundances. We use Eq. \ref{pil-cal}
above to estimate the oxygen abundances for this galaxy.

We estimated the mean oxygen abundance parameter, $R_{23}$, by summing
the fluxes of the nebular emission lines at different regions along
the ring-like structure of VGS31b (see Table \ref{fluxes}). Such
regions have been chosen by accounting the results of the 1D model of
the light profile along the ring major axis, discussed in Sec
\ref{mag}: the ring is the dominant component for $r \geq 5$, but,
taking into account that the HG effective radius is $r_{e} \sim 6.8$
arcsec, as derived by the 1D model, in order to avoid regions in which
there was the ``contamination'' of the HG, we measured the line fluxes
in the range $0<r<7$ arcsec for the host galaxy and $r > 7$ arcsec for
the ring, obtaining a value of the oxygen abundance of
$12+log(O/H)_{P} = 8.6 \pm 0.6$ for the HG and $12+log(O/H)_{P} = 8.37
\pm 0.57$ for the ring. In Table \ref{ox} we show the comparison
between the average value of oxygen abundance obtained for VGS31b and
those obtained for other PRGs, NGC4650A \citep{Spav10}, UGC7576 and
UGC9796 \citep{Spav11}. \\ Given the ``irregular'' morphology of the
ring in VGS31b, due to the presence of several ``holes'' inside its
extension (see Fig.\ref{VGS31b}), contrary to the well-defined
disk-like morphology in NGC4650A, for instance, it was difficult to
derive the oxygen abundance in the ring of VGS31b as a function of the
radius. Thus, the available measurements let us to check that the
metallicity remains almost constant along the projected major axis of
the ring, except for a slight increase towards the center. The same
behaviour has been observed for other PRGs \citep{Spav10, Spav11},
unlike what it is observed for spiral galaxies, where there is a very
steep gradient in metallicity.

By assuming the oxygen abundance and metallicity for the Sun to be 
$12 + log(O/H)_{\odot} = 8.83 \pm 0.20=
A_{\odot}$ and $Z_{\odot} = 0.02$ \citep{Gre98}, given that $Z \approx
K Z_{\odot}$, $K_{VGS31b} = 10^{[A_{VGS31b} - A_{\odot}]}$, we obtain
a metallicity for the ring-like structure of $Z \simeq (0.35 \pm 0.03)
Z_{\odot}$ (see Table \ref{ox}).\\ The main implications of this analysis
will be described and discussed in the following sections.

\begin{table*}
\caption{\label{fluxes}Observed and de-reddened mean emission line fluxes relative to $H \beta$.} \centering
\begin{tabular}{lccc}
\hline\hline
line& $\lambda$ (\AA) & Observed flux relative to $H \beta$ &De-reddened flux relative to $H \beta$\\
\hline
$[OII]$ & 3727 & 4.9 &1.2\\
$[OIII]$ & 4959 & 2.1 & 2.4\\
$[OIII]$ & 5007 & 5.9 & 7.1\\
$H \gamma$ & 4340 & 1.2 & 0.6\\
$H \alpha$ & 6563 & 0.7 & 2.7\\
\hline
\end{tabular}
\end{table*}

\begin{table*}
\caption{\label{ox}Oxygen abundances and metallicities of VGS31b compared to those obtained for UGC7576, UGC9796 and NGC4650A.} \centering
\begin{tabular}{lcccc}
\hline\hline
Parameter&VGS31b&UGC7576&UGC9796&NGC4650A\\
\hline
$12+log(O/H)$ & $8.37 \pm 0.57 $ & $8.5 \pm 0.5 $ & $7.7 \pm 1 $ & $8.2 \pm 0.2$\\
$Z/Z_{\odot}$ & $0.30 \pm 0.03 $ & $0.4 \pm 0.01 $ & $0.1 \pm 0.005$ & $0.2 \pm 0.002$\\
\hline
\end{tabular}
\end{table*}

\section{Metallicity-luminosity relation}\label{ML}
\begin{figure*}
\centering
\includegraphics[width=12cm]{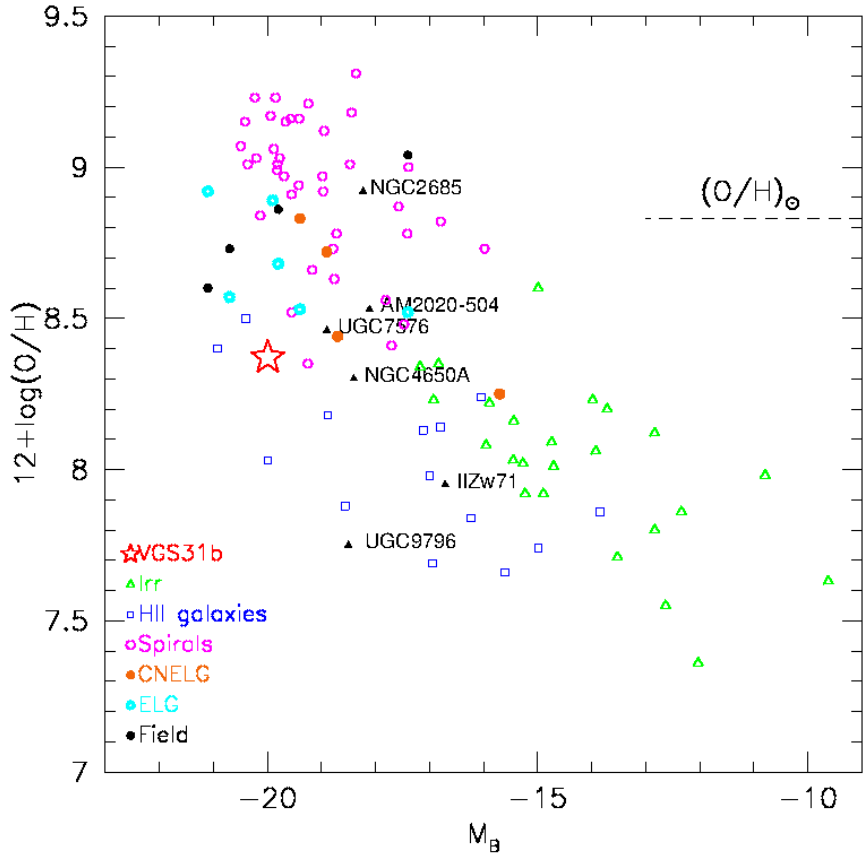}
\caption{Oxygen abundance vs absolute blue magnitude for Compact
  Narrow Emission-Line Galaxies (CNELGs, orange filled circles),
  star-forming Emission Line Galaxies (ELGs, cyan open circles), four
  field galaxies with emission lines (filled black circles), nearby
  dwarf irregulars (open triangles), local spiral galaxies (open
  circles), local HII galaxies (open squares). The sample of late-type
  disk galaxies are by \citet{Kob99}. PRGs are: VGS31b (red star),
  NGC4650A \citep{Spav10}, IIZw71 \citep{Per09}, NGC2685
  \citep{Esk97}, AM2020-504 \citep{Frei12}, UGC7576 and UGC9796
  \citep{Spav11}. The dashed line indicates the solar oxygen
  abundance.} \label{conf}
\end{figure*}

The mean value for the oxygen abundance along the ring-like structure
of VGS31b, derived by the empirical method (see sec. \ref{oxy}), as a
function of the total luminosity, is compared with those for different
late-type galaxies by \cite{Kob99}\footnote{The absolute blue
  magnitude for the objects in the sample of \citet{Kob99} are
  converted by using $H_{0} =75$ km/s/Mpc.} (spirals, Irregulars and
HII galaxies) and several PRGs: results are shown in
Fig. \ref{conf}.

We found that, in the metallicity-luminosity relation, VGS31b has
lower metallicity with respect to those typical for spiral galaxies of
comparable total luminosity (i.e. $M_B \sim -20$), and it is located
in the region where some other PRGs are found, in particular the polar
disk NGC4650A \citep{Spav10} and the polar ring UGC7576
\citep{Spav11}. Taking the total magnitude into account, such value
of oxygen abundance is somewhat lower than expected by the
metallicity-luminosity relation.

\section{Star formation rate}\label{SFR}

The $H \alpha$ emission is detected with adequate signal-to-noise,
and from the measured integrated flux we can derive the Star Formation
Rate (SFR) for the ring-like structure of VGS31b: from the
$H \alpha$ luminosity, using the expression given by \cite{Ken98},
we estimate a $SFR = 7.9 \times 10^{-42} \times L(H \alpha)$. We
find that it is almost constant along the ring, within a large scatter
in the individual values. From the average value of $L(H \alpha)
\simeq 2.5 \times 10^{39}$ erg/s we have obtained an average $SFR \sim
0.02 M_{\odot}/yr$.

By using the {\it g-r} integrated color and adopting the stellar
population synthesis model by \citet{Bru03}, described in
Sec.\ref{mag}, we estimated an upper limit for the age of VGS31b of
1Gyr. This suggests that the ring-like structure in VGS31b is even
younger. Given that the last burst of star formation occurred less
than 1 Gyr ago, we also checked if the present SFR and even 2 and 3
times higher (i.e. $SFR \sim 0.02 M_{\odot}/yr$, $SFR \sim 0.04
M_{\odot}/yr$ and $SFR \sim 0.06 M_{\odot}/yr$) can give the inferred
metallicities of $Z=0.3 Z_{\odot}$.

We used a linearly declining SFR \citep{Bru03} $\psi(t) =
2M_{\star}\tau^{-1}[1-(t/\tau)]$ (typically used for late-type
galaxies), to estimate the expected stellar mass for the three
different values of the SFR and three epochs (0.8 Gyr, 1 Gyr and 2
Gyrs), obtaining stellar masses in the range $1.3 \times 10^{8}
M_{\odot} \leq M_{\star} \leq 4.3 \times 10^{8} M_{\odot}$. Than, by
using the mass-metallicity relation derived by \citet{Tre04}, where
$12+log(O/H) = -1.492 + 1.847 log(M_{\star}) - 0.08026 (log
M_{\star})^{2}$, we found that $0.2 Z_{\odot} \le Z \le 0.5
Z_{\odot}$. In particular, the present SFR for the ring-like
structure is able to increase the metallicity of about 0.05
$Z_{\odot}$ after 1Gyr. We found that the metallicity of $Z=0.30 \pm
0.03 Z_{\odot}$, estimated by using the element abundances, falls near
the lower limit of the range of expected metallicities. The
implications of these results will be discussed in detail in
Sec. \ref{conc}.

\section{Discussion and conclusions} \label{conc}

VGS31b is a peculiar galaxy in a small group of three interacting
galaxies, all embedded in a unique HI envelope, located in a void
region, whose properties are very recently analysed by
\citet{Beygu13}. They showed that the HI is distributed as a
``filamentary structure'' and some gas is accreted by the galaxies
from the filament. VGS31b has a ring-like structure and a tail
detected both in the optical and in the HI images. This object, as
already suggested by \citet{Beygu13}, can be a good candidate for a
ring formation through the cold accretion of gas along a filament, as
other few cases already known \citep{Kreckel12, Sta09}.

In this work, we have presented a detailed photometric and
spectroscopic study of this object, based on data extracted from SDSS
and WHT archives and the main goal is to trace the possible formation
history for this object. In particular, we aim to test the cold
accretion of gas through a ``cosmic filament'' as a possible scenario
for the formation of the ring-like structure in this galaxy. In the
following, we address how the results obtained for VGS31b can be
reconciled with the theoretical predictions for the proposed scenarios
for the formation of PRGs and related objects.

The main results of the analysis performed in the present paper are:
\begin{enumerate}
\item the structure of VGS31b, as appears by the optical images (see
  Sec.\ref{morph}) resembles that of forming highly-inclined ring
  galaxy: most of the light comes from the central spheroidal
  component, while the ring-like structure, inclined of about 70
  degrees from the HG equatorial plane, and the tail, at the
  North-East side, become very faint towards longer wavelengths,
  suggesting a very young stellar population for this component;

\item the ring seems to reach the galaxy center tracing a
  ``spiral-like'' pattern, where the ring's arms are connected to the
  two bright blobs at the ends of the bar-like structure and to the
  two inner arcs;

\item optical g-r and g-i colors show that the inner regions of the HG
  are characterised by several areas of very blue colors (the nucleus,
  inside 4 arcsec from the center, two knots along the HG major axis
  and one more on the East side with respect to the galaxy nucleus),
  which suggest the presence of star forming regions, while both the
  ring-like structure and the tail lack of these features: these turn
  to be consistent with the results previously found by
  \citet{Beygu13};

\item The ring-like structure of VGS31b has an average metallicity of
  $Z=0.3 Z_{\odot}$, which is lower with respect to that of
  same-luminosity spiral disks, but it turns to be consistent with the
  values derived for other PRGs. This value remains almost constant
  along the whole extension of the ring;

\item The tail at the North-East side has integrated magnitudes
  corresponding to stellar masses of the order of $10^{8} M_{\odot}$.
\end{enumerate}

The study of both the chemical abundances of HII regions in polar ring
galaxies and their implications for the evolutionary scenario of these
systems has been a step forward both in tracing the formation history
of the galaxy and giving hints toward the mechanisms at work during
the building of a disk by cold accretion process. To account both for
the featureless morphology of the central spheroidal galaxy and for
the more complex structure of the polar ring/disk, three main
formation processes have been proposed so far: {\it i)} a major
dissipative merger (\citealt{Bek97}; \citealt{Bek98a};
\citealt{Bou05}), {\it ii)} tidal accretion of material (gas and/or
stars) from outside (\citealt{Res97}; \citealt{Bou03};
\citealt{Han09}), and {\it iii)} cold accretion of pristine gas along
a filament \citep{Mac06, Bro08, Snaith12}.

As suggested by the previous studies of PRGs
(\citealt{Spav10,Spav11,Spav12}, \citealt{Iod06}), the critical
physical parameters that can help to distinguish among the three
formation scenarios are 1) the total baryonic mass (stars plus gas)
observed in the polar structure with respect to the central spheroid,
2) the kinematics along both the equatorial and meridian planes, 3)
the metallicity and SFR in the polar structure.\\ If the polar
structure around both an elliptical and disk galaxy forms by the cold
accretion of gas from filaments there is no limit to the accreted
mass. Moreover, due to the inflow of pristine gas, the metallicity is
lower than that observed in galaxies of a comparable total luminosity,
and the value derived by the present SFR is higher than those directly
measured by the chemical abundances. \cite{Spav10,Spav11} find that
such predictions are consistent with observations in the polar disk
galaxy NGC4650A and in the polar ring galaxy UGC7576, leading to the
conclusion that the cold accretion of gas by cosmic web filaments is
the most realistic formation scenario for these objects. This
observational results have been confirmed by recent simulations
\citep{Snaith12}.

 The tidal accretion scenario, in which gas is stripped from a
 gas-rich donor in a particular orbital configuration \citep{Bou03},
 is able to produce polar rings and/or disks both around a disk or an
 elliptical galaxy. VGS31 is a small group of three interacting
 gas-rich galaxies in a void, thus, this scenario could be the
 favourite one and the close companion VGS31a could be the donor
 galaxy.  An important constraint that need to be considered to
 exclude or not this scenario is the distribution of the HI gas: the
 three galaxies share the same envelope and they are almost aligned
 along a narrow filament. Thus, if the ring has formed by the tidal
 interaction between VSG31b and VSG31a, this observed configuration
 can also settled after that event, otherwise the HI distribution
 should be destroyed during the first galaxy encounter. Moreover, as
 already suggested by \citet{Beygu13}, as the mass ratio between
 VSG31b and VSG31a is 3 to 1, a tidal interaction would damage the
 disks and produce more prominent tails.

Moreover, we analysed the possibility that the disruption of a dwarf
galaxy could account for the formation of the ring-like structure and
the tail observed in VGS31b. As discussed in section \ref{mag}, we
used stellar population synthesis models \citep{Bru03} to estimate the
stellar mass corresponding to the integrated magnitude observed in the
tail of VGS31b. We derived a stellar mass of the order of $10^{8}
M_{\odot}$ for the tail, which turns to be higher than the typical
masses of dwarf galaxies ($10^{3} M_{\odot}\leq M_{\ast} \leq 10^{7}
M_{\odot}$, \citealt{Saw11}).

Can the ring-like structure observed in VGS31b be the result of a
polar major merging between two disk galaxies already present in the
filament? According to \cite{Bou03}, to form a ring with a diameter of
about 24 kpc, as observed in this object (see Sec.\ref{vgs31}), the
relative velocities between the two galaxies should be not larger than
120 km/s and the stellar mass of the victim disk, which will form the
ring, should be of the order of $10^{10} M_\odot$. Since we do not
have information about the orbit of the collision, and about relative
velocities, the key parameter for this scenario is the stellar mass of
the remnant victim galaxy, i.e. the ring: the value estimated by the
integrated colors is about $4 \times 10^{8} M_\odot$, which is two
order of magnitude lower than that requested in the simulation to form
an extended ring as observed in VGS31b.

As already suggested by \citet{Beygu13}, the large HI amount and, most
of all, its distribution let VGS31b to be a good candidate for a ring
formation through the cold accretion of gas along the filament where
this galaxy and its companions are embedded in. The new result of the
present work, which reinforces this view, is the low metallicity
measured for the ring in VGS31b, i.e. $Z=0.3Z_{\odot}$, that remains
almost constant along the whole extension (see Sec.\ref{oxy}): this
turns to be consistent with the predictions by recent simulations of
\citet{Snaith12}, where they found that the typical metallicity for a
polar disk formed through the cold accretion of gas along a filament
is about $Z=0.2Z_{\odot}$ and any significant gradient is measured
along the polar structure.

Given all the evidence shown above, we can conclude that the cold
accretion of gas by the filament could well account for the main
properties of VGS31b. Anyway, to check this scenario for the ring
formation in VSG31b and exclude the other ones, one needs an ad-hoc
simulation, that should reproduce both the observed properties of this
object and, at the same time, the global structure observed for the
whole group of galaxies.  In particular, in all the simulations
proposed for the PRGs formation by the cold accretion \citep{Mac06, Bro08, Snaith12}, the polar structure has the characteristic of a
disk, rather than a ring: is the highly-inclined structure observed in
VGS31b a disk still forming? Observations show that the tail, optical
and HI, is kinematically associated to the ring and this reaches the
galaxy center by tracing a ``spiral-like'' pattern (see
Sec.\ref{morph}): these evidences could be consistent with an
highly-inclined ring structure which is still forming, and, according
to the simulations by \citet{Bro08}, it could be in the phase of the
last merging through the filament, when the angular-momentum
decoupling is happening.

\section*{Acknowledgements}
We wish to thank the anonymous referee for helpful comments, which
allow us to improve the paper. This paper makes use of data obtained
from the Isaac Newton Group Archive which is maintained as part of the
CASU Astronomical Data Centre at the Institute of Astronomy,
Cambridge. M.S. acknowledge financial contribution from the ``Fondi di
Ateneo 2011'' (ex 60 \%) of Padua University.

\bibliography{bibliografia.bib}

\bsp

\label{lastpage}

\end{document}